\pdfoutput=1
\documentclass[11pt,a4paper]{article}

\usepackage[T1]{fontenc}
\usepackage[utf8]{inputenc}
\usepackage{XCharter}                       % text face: Charter
\usepackage[xcharter,vvarbb]{newtxmath}     % matching math
\usepackage[scale=0.92]{sourcesanspro}      % headings, captions
\usepackage[varqu,scaled=0.92]{zi4}         % monospace (URLs)
\usepackage{textcomp}
\usepackage{microtype}

\usepackage[a4paper,top=26mm,bottom=26mm,left=27mm,right=27mm,
            headheight=14pt,headsep=8mm,footskip=11mm]{geometry}
\usepackage{amsmath}
\usepackage{siunitx}
\usepackage[version=4]{mhchem}

\usepackage{graphicx}
\usepackage{booktabs}
\usepackage{xcolor}
\definecolor{accent}{RGB}{18,72,140}
\definecolor{softgray}{gray}{0.45}
\usepackage[font=small,labelfont={bf,sf,color=accent},labelsep=period,
            justification=justified,singlelinecheck=true,
            skip=6pt]{caption}
\usepackage{enumitem}
\setlist[itemize]{leftmargin=1.6em,itemsep=2pt,topsep=4pt}

\usepackage{titlesec}
\titleformat{\section}
  {\sffamily\bfseries\Large}{\textcolor{accent}{\thesection}}{0.75em}{}
\titleformat{\subsection}
  {\sffamily\bfseries\large}{\textcolor{accent}{\thesubsection}}{0.7em}{}
\titlespacing*{\section}{0pt}{3.0ex plus 1ex minus .3ex}{1.2ex plus .3ex}
\titlespacing*{\subsection}{0pt}{2.4ex plus 1ex minus .3ex}{0.9ex plus .2ex}

\usepackage{fancyhdr}
\fancypagestyle{firstpage}{%
  \fancyhf{}%
  \fancyfoot[C]{\footnotesize\sffamily\thepage}%
  }

\usepackage[numbers,square,sort&compress]{natbib}
\usepackage[hyphens]{url}
\usepackage[colorlinks=true,linkcolor=accent,citecolor=accent,
            urlcolor=accent,pdfstartview=FitH,bookmarksnumbered=true,
            hyperfootnotes=false]{hyperref}
\usepackage{orcidlink}
\hypersetup{
  pdftitle={Edge-scaling softening of a universal machine-learned potential:
            point-defect formation energies in a coronene graphene quantum dot},
  pdfauthor={Andrii Tuzhykov, Vladimir Soloviev, Arnold Kiv, Andrii Bielinskyi},
  pdfsubject={Benchmark of a universal machine-learning interatomic potential
              (SevenNet) against spin-polarized DFT for point defects in coronene},
  pdfkeywords={machine-learning interatomic potential, surrogate model,
               graphene quantum dot, coronene, edge defect, formation energy,
               softening bias, model validation, SevenNet,
               density functional theory}}
\newcommand{\doi}[1]{\href{https://doi.org/#1}{doi:\nolinkurl{#1}}}
\newcommand{\coronene}{\ce{C24H12}}
\newcommand{\Dsixh}{\ensuremath{D_{6h}}}
\newcommand{\Ef}{\ensuremath{\Delta E_f}}
\newcommand{\eVA}{eV\,\AA$^{-1}$}

\begin{document}
\thispagestyle{firstpage}

% ---------------------------- Title block ----------------------------
\begin{center}
  {\sffamily\bfseries\fontsize{15.5}{20}\selectfont
   Edge-scaling softening of a universal machine-learned potential:\\[1pt]
   point-defect formation energies in a coronene graphene quantum dot\par}
  \vspace{1.3em}
  {\large
   \mbox{Andrii Tuzhykov\textsuperscript{1,$\ast$,$\dagger$}\,\orcidlink{0009-0005-3853-9911}},\hspace{1.2em}
   \mbox{Vladimir Soloviev\textsuperscript{1,3,$\dagger$}\,\orcidlink{0000-0002-4945-202X}},\\[3pt]
   \mbox{Arnold Kiv\textsuperscript{1,2,$\dagger$}\,\orcidlink{0000-0002-0991-2343}},\hspace{1.2em}
   \mbox{Andrii Bielinskyi\textsuperscript{3,4,$\dagger$}\,\orcidlink{0000-0002-2821-2895}}\par}
  \vspace{0.9em}
  \newcommand{\affil}[3]{%
    {\small\textsuperscript{#1}\,#2}\\
    {\footnotesize\itshape\color{softgray}#3}\par\vspace{2.5pt}}
  \affil{1}{South Ukrainian National Pedagogical University named after K.\,D.~Ushynsky}
        {Staroportofrankivska Street 26, 65020 Odesa, Ukraine}
  \affil{2}{Ben-Gurion University of the Negev}
        {David Ben-Gurion Blvd.~1, 8410501 Beer-Sheva, Israel}
  \affil{3}{Kyiv National Economic University named after Vadym Hetman}
        {Beresteysky Prospect 54/1, 03057 Kyiv, Ukraine}
  \affil{4}{State University of Economics and Technology}
        {Medychna Street 16, 50005 Kryvyi Rih, Ukraine}
\end{center}

% First-page notes (unnumbered footnote block)
\begingroup
\makeatletter
\renewcommand{\@makefntext}[1]{\noindent\raggedright#1}
\makeatother
\renewcommand{\thefootnote}{}
\footnotetext{%
  \textsuperscript{$\ast$}\,Corresponding author:
  \mbox{\href{mailto:andrewtuzhykov@gmail.com}{andrewtuzhykov@gmail.com}} (A.~Tuzhykov).\quad
  \textsuperscript{$\dagger$}\,These authors contributed equally.\\
  E-mail addresses:
  \mbox{\href{mailto:vnsoloviev2016@gmail.com}{vnsoloviev2016@gmail.com}} (V.~Soloviev);
  \mbox{\href{mailto:kiv.arnold20@gmail.com}{kiv.arnold20@gmail.com}} (A.~Kiv);
  \mbox{\href{mailto:bielinskyi.andrii99@gmail.com}{bielinskyi.andrii99@gmail.com}} (A.~Bielinskyi).\par
  \vspace{3pt}
  Presented at the ADT-ISMDDC-2026 Workshop ``Advanced Digital Technologies: Intelligent Systems,
  Modeling, and Data-Driven Control'', co-located with ICST-ODESA-2026: XIV International Scientific
  and Practical Conference ``Information Control Systems and Technologies'', September 22--24, 2026,
  Odesa, Ukraine. To appear in \emph{CEUR Workshop Proceedings}.\par
  \vspace{3pt}
  \textcopyright\,2026 Copyright for this paper by its authors. Use permitted under Creative Commons
  License Attribution 4.0 International
  (\href{https://creativecommons.org/licenses/by/4.0/}{CC BY 4.0}).}
\endgroup

% ------------------------------ Abstract -----------------------------
\vspace{0.4em}
\noindent\begin{minipage}{\textwidth}
{\color{black!20}\rule{\textwidth}{0.4pt}}\par\vspace{0.9em}
\centering
\begin{minipage}{0.94\textwidth}
\small
{\sffamily\bfseries\color{accent}Abstract}\par\smallskip
\noindent
Universal machine-learning interatomic potentials (MLIPs) are increasingly used as computationally
efficient surrogates for density functional theory (DFT) in high-throughput nanomaterials screening.
Systematic softening has been reported for several universal MLIPs when high-energy and
far-from-equilibrium configurations are underrepresented in training data. We examine how this
behavior appears at under-coordinated edge sites in graphene quantum dots (GQDs), where a vacancy can
have pronounced molecular-radical character. Using coronene (\coronene), a hexagonal GQD with \Dsixh{}
symmetry and three carbon classes, we benchmark SevenNet (7net-omni-i12) against spin-polarized PBE
DFT (GPAW) for monovacancy, divacancy, and Stone--Wales defects. Beyond the full site- and
bond-resolved landscapes, we perform a controlled matched-reconstruction comparison in which 28
vacancy and divacancy configurations are scored by both engines using self-consistent
chemical-potential reservoirs. In this dataset, SevenNet underestimates all 28 vacancy and divacancy
formation energies, and the mean deficit increases with under-coordination within each defect
family: 0.30--0.74~eV for monovacancies and 0.65--1.11~eV for divacancies as the number of broken
edge C--H bonds increases. By contrast, the mean Stone--Wales energies differ by only 0.013~eV.
SevenNet also preserves the ordering of three hydrogen-dependent reconstruction basins and
reproduces \Dsixh-equivalent sites to within a few meV. The within-class symmetry spread therefore
provides a practical diagnostic of basin-sensitive relaxations. Within the present coronene
benchmark, these results support MLIP-based structure screening and preliminary ranking, while
absolute vacancy energies and multi-basin edge configurations still require DFT validation.

\par\medskip\noindent
{\sffamily\bfseries\color{accent}Keywords:}\enspace
machine-learning interatomic potential, surrogate model, graphene quantum dot, coronene, edge
defect, formation energy, softening bias, model validation, SevenNet, density functional theory
\end{minipage}\par\vspace{0.9em}
{\color{black!20}\rule{\textwidth}{0.4pt}}
\end{minipage}
\vspace{0.4em}

% =====================================================================
\section{Introduction}\label{sec:intro}

Graphene quantum dots (GQDs) are finite, typically nanometer-scale graphene fragments whose optical,
magnetic, catalytic, sensing, and electrode properties are governed jointly by quantum confinement
and edge chemistry~\cite{geng2026}. Point defects, including monovacancies, divacancies, and
Stone--Wales rotations, reorganize the local $\pi$ network and can create under-coordinated
open-shell sites. Reliable site-resolved formation energies are therefore required for predictive
GQD design.

Spin-polarized density functional theory (DFT) provides the reference description but becomes
costly when every symmetry-inequivalent site and reconstruction must be relaxed. Universal MLIPs
based on equivariant graph neural networks, including NequIP~\cite{batzner2022},
MACE~\cite{batatia2022}, and SevenNet~\cite{park2024}, enable high-throughput screening at
substantially lower computational cost. A documented limitation is systematic softening: training
sets dominated by relaxed structures can lead to underestimated energies for far-from-equilibrium
configurations, including defects~\cite{deng2025}. The error may be particularly pronounced at
hydrogen-terminated GQD edges, where vacancy formation can break a C--H bond and create radical
environments poorly represented in bulk-dominated training data.

A rigorous benchmark must also control for competing reconstruction basins. Edge radicals can have
near-degenerate, multireference states, so unconstrained DFT and MLIP optimizations may converge to
different local structures~\cite{nieman2017}. Formation energies additionally depend on the
chemical-potential reservoirs. We therefore use coronene (\coronene), whose \Dsixh{} symmetry
reduces 24 carbon sites to three classes, to compare monovacancy, divacancy, and Stone--Wales
landscapes obtained with SevenNet and spin-polarized PBE DFT. For 28 vacancy and divacancy
configurations, the same relaxed geometry is evaluated with both methods using method-consistent
reservoirs. This design separates model-dependent energy differences from basin mismatch and tests
whether the error grows with edge under-coordination.

% =====================================================================
\section{Background and related work}\label{sec:background}

\subsection{Defects in graphene and graphene quantum dots as a problem of local chemistry and
energetics}\label{sec:bg-defects}

Point defects in graphene modify local coordination, $\pi$-electron density, strain, magnetism, and
chemical reactivity. Monovacancies commonly reconstruct into a 5-9 motif with one undercoordinated
carbon, whereas divacancies adopt 5-8-5, 555-777, or 5555-6-7777 motifs that redistribute
undercoordination and strain. A Stone--Wales rotation preserves atom count but forms a 5-7/5-7 pair
and alters local curvature and reactivity~\cite{skowron2015,ma2004,lee2005,ma2009,tiwari2023}.
These structural differences affect catalytic, adsorption, transport, optical, and spin
properties~\cite{skowron2015,ma2004,ma2009,tiwari2023}.

However, most of the classical literature on defects in graphene was developed for infinite or
periodic models, where the defect energy is treated as a bulk-like quantity that converges with
supercell size. The situation is different in graphene quantum dots (GQDs) and finite polycyclic
aromatic hydrocarbons (PAHs): the edge is itself a chemically distinct object, rather than a weak
perturbation. In small hydrogen-terminated fragments such as coronene, \coronene, the internal
graphene-like region and the periphery coexist within the same molecule. Therefore, the defect
formation energy is governed not only by breaking C--C bonds in the $\pi$ network, but also by local
edge passivation, the presence or absence of C--H bonds, possible radical stabilization, and
competition among several nearby minima on the potential energy
surface~\cite{santana2013,wagner2013,vanin2010,gao2021,cho2024,martin2024,forte2008}. For this
reason, monovacancies, divacancies, and Stone--Wales defects in coronene should not be treated as
simple miniature analogues of the same defects in periodic graphene.

Two literature traditions are relevant to the present problem. Periodic DFT studies emphasize
supercell convergence, spin polarization, and defect-core reconstruction, whereas molecular PAH,
nanographene, and GQD studies emphasize discrete electronic states, open-shell character, edge
passivation, and size effects. For coronene, both perspectives are needed: the molecular framework
captures finite-edge chemistry, while much of the quantitative defect-energy baseline comes from
periodic
calculations~\cite{nieman2017,skowron2015,martin2024,forte2008,thomas2022,machado2015,pinheiro2019,nieman2023,shtepliuk2018,lin2015}.

\subsection{DFT energetics of monovacancies, divacancies, and Stone--Wales defects in
graphene}\label{sec:bg-dft}

The most reliable baseline for a monovacancy in monolayer graphene is provided by spin-polarized DFT
calculations, which yield formation energies of approximately 7.3--8.1~eV depending on the
functional, supercell size, and relaxation protocol. In the review by Skowron et al., periodic
PBE/PW91/LDA calculations give a broad but consistent range of approximately 7.4--7.9~eV. Early
calculations by Lee et al.\ and Ma et al.\ established the reconstructed 5-9 geometry and the
presence of a magnetic defect center, while later multi-method studies emphasized that the magnetic
moment and the planarity or nonplanarity of the local geometry are sensitive to the exchange
component of the functional and to the choice of a cluster or periodic
setup~\cite{skowron2015,ma2004,lee2005,valencia2017}. The characteristic migration barriers for a
monovacancy in graphene are approximately 1.1--1.4~eV, which is low enough for thermally activated
mobility to be relevant already in a moderate temperature range~\cite{skowron2015,lee2005}.

The DFT literature on divacancies is richer in terms of reconstruction motifs. The most frequently
discussed structures are V$_2$(5-8-5), V$_2$(555-777), and V$_2$(5555-6-7777). According to the
survey by Skowron et al., the formation energy of V$_2$(5-8-5) in infinite graphene is usually in
the range of 7.5--8.0~eV, whereas conversion to V$_2$(555-777) provides an additional stabilization
of approximately 0.8--1.0~eV. Thus, the 555-777 structure is generally the lower-energy
reconstruction relative to 5-8-5. Separate calculations by Kim et al.\ give an energy difference of
approximately 0.93~eV between 5-8-5 and 555-777, with reconstruction barriers between these states
of approximately 5.1--5.3~eV, explaining their metastability at room
temperature~\cite{skowron2015,kim2011}. Earlier work on divacancies in graphene and carbon nanotubes
also emphasized that the relative stability of 5-8-5 and 555-777 is controlled by the balance
between removal of undercoordination and local elastic distortion~\cite{amorim2007}.

The Stone--Wales defect occupies a special position in this series. Unlike vacancies, it does not
change the atomic composition and preserves the three-coordinated network, although it creates a
pronounced topological and elastic anomaly. In graphene, its formation energy is typically around
5~eV; values of approximately 5.0--5.3~eV are commonly cited, and some calculations give about
5.08~eV for sufficiently large supercells~\cite{ma2009,tiwari2023,lusk2008}. This energy is lower
than that of a monovacancy, but the defect is associated with high barriers for both formation and
reverse relaxation, making it kinetically stable once formed~\cite{ma2009,tiwari2023}. The
coordination-preserving nature of the Stone--Wales rearrangement is also important for interpreting
the behavior of universal machine-learning interatomic potentials (MLIPs).

An important refinement of this picture was provided by diffusion quantum Monte Carlo calculations.
Thomas et al.\ showed that standard DFT approaches underestimate the monovacancy formation energy in
graphene by approximately 1~eV on average, whereas the discrepancy is noticeably smaller for
Stone--Wales defects~\cite{thomas2022}. This observation is important here in two respects. First,
it shows that even within benchmark quantum-chemical literature, absolute defect energies are not
universal constants. Second, it indicates that the largest errors arise precisely for the most
locally disrupted and undercoordinated configurations, whereas coordination-preserving topological
rearrangements are usually described more robustly~\cite{ma2009,thomas2022}.

Berger, Bagheri, and Komsa~\cite{berger2025} demonstrated high-throughput vacancy screening with
universal MLIPs. Benchmarking MACE, CHGNet, M3GNet, and ALIGNN against DFT defect datasets, they
reported typical MACE vacancy-formation-energy RMSEs of about 0.4--0.8~eV and then screened 86,259
Materials Project structures. Their results support MLIP-assisted screening while also emphasizing
limitations for more complex defects, small energy differences, and defective configurations that
are underrepresented in training data.

Cross-study comparisons require caution because formation energies depend on the
exchange-correlation functional, spin treatment, supercell size, boundary conditions, relaxation
protocol, chemical-potential reservoirs, and system type~\cite{skowron2015,thomas2022,valencia2017}.
It is therefore preferable to compare ranges and systematic trends rather than a single reference
value. Across the cited literature, monovacancies generally cost more than Stone--Wales defects, the
555-777 reconstruction commonly stabilizes divacancies relative to 5-8-5, and finite-size and edge
effects can substantially modify peripheral-defect
energetics~\cite{skowron2015,santana2013,thomas2022}.

\subsection{Edge, passivation, and finite size as determining factors in defect
energetics}\label{sec:bg-edge}

One of the most important results for finite graphene fragments is that edge defects are not simply
``weakened'' internal defects. Santana, Popov, and Bichoutskaia showed for small graphene flakes
that a monovacancy at the edge is thermodynamically much more favorable than one at the center. The
energy difference between a flake with a central vacancy and one with an edge vacancy is
6.6--8.9~eV for the considered sizes; in the C$_{116}$ model, stepwise displacement of the vacancy
toward the edge decreases this difference from 8.0~eV for a central position to almost zero at the
periphery~\cite{santana2013}. They also found one-step migration barriers of ${\sim}0.57$~eV toward
the edge and ${\sim}0.75$~eV away from the edge, indicating not only a static but also a kinetic
tendency toward the periphery~\cite{santana2013}. This is particularly relevant to coronene: in such
a small GQD, central and edge sites are separated by only a few rings, so bulk-like and edge-like
chemistry inevitably mix.

The physical origin of this edge stabilization is multicomponent. In the interior, a monovacancy
creates a more rigid combination of local compression, undercoordination, and disruption of $\pi$
conjugation. At the edge, part of this cost can be relieved by peripheral reconstruction, changes in
the number of dangling bonds, and local rearrangement of the already existing edge geometry. For
this reason, graphene edge reconstructions are often described as self-passivating. The work of
Koskinen, Malola, and H\"akkinen on self-passivating edge reconstructions, Rodrigues et al.\ on
zigzag-edge reconstruction through Stone--Wales motifs, and Kroes et al.\ on the mechanism and
free-energy barrier of the 57 reconstruction all show that an edge can reduce the number of active
undercoordinated centers by forming non-hexagonal
motifs~\cite{koskinen2008,rodrigues2011,kroes2011}. Experimentally, this is consistent with the
observations by Kim et al.\ of atomically perfect torn edges and their reversible
reconstruction~\cite{kim2013}. Thus, the formation energy of an edge defect depends not only on how
many atoms are removed, but also on what new edge motif is formed after relaxation.

Edge geometry and chemistry are further complicated by hydrogen termination. For H-terminated edges,
it is not always sufficient to distinguish only between armchair and zigzag motifs. The DFT study by
Wagner et al.\ showed that fully hydrogenated Klein and reconstructed-Klein edges can be
energetically more favorable than the conventionally discussed fully hydrogenated zigzag
edges~\cite{wagner2013}. Vanin et al.\ calculated the formation free energies of different edge
configurations in \ce{H2}, \ce{O2}, \ce{N2}, \ce{CO}, \ce{CO2}, and \ce{H2O} atmospheres,
demonstrating a strong dependence of edge stability on chemical environment~\cite{vanin2010}. A
later study by Gao et al.\ separately confirmed that hydrogenation stabilizes the edge by
compensating dangling bonds through C--H bond formation~\cite{gao2021}. For an H-terminated GQD,
this means that removing an edge carbon atom is not equivalent to removing an internal atom: in
general, both the C--C framework and the local C--H passivation regime are disrupted, and the final
state may contain a different set of radical and/or self-passivated motifs. This feature is
especially critical for coronene, where peripheral atoms constitute a large fraction of all atoms in
the system~\cite{wagner2013,vanin2010,gao2021}.

Edge functionalization more generally alters charge distribution, local sp$^2$/sp$^3$ character,
and structural deformation~\cite{vanin2010,cho2024}. It can therefore change defect costs by
passivating undercoordinated atoms or by introducing additional bonds and chemical-potential terms
that must be included in the formation-energy definition. For the present H-terminated system, this
broader literature mainly motivates treating hydrogen chemistry explicitly rather than assuming that
an edge vacancy is determined solely by disruption of the carbon
framework~\cite{vanin2010,gao2021,cho2024}.

Finite size is equally important. In a molecular approach, Forte et al.\ considered a series of
clusters ranging from pyrene to coronene, coronene-19, coronene-37, and coronene-61. They showed
that in pristine clusters the HOMO--LUMO gap decreases with increasing size, whereas in the presence
of a vacancy the gap does not decrease as strongly and the local magnetic and electronic response
persists~\cite{forte2008}. In other words, even when a central defect structurally resembles a
bulk-graphene vacancy, electronically it remains a molecular defect over a substantial size range.
Santana et al.\ add a geometric scale to this picture: in their models, directed vacancy motion
toward the edge begins to dominate at distances to the edge of approximately
10~\AA~\cite{santana2013}. Consequently, in dots at the scale of coronene and even circumcoronene, a
central defect does not have to be fully bulk-like; its energy may be noticeably affected by the
periphery even when the defect is not located directly at the
rim~\cite{santana2013,martin2024,forte2008}.

\subsection{Machine-learning interatomic potentials and their limitations for defective
configurations}\label{sec:bg-mlip}

MLIPs have expanded defect simulations beyond classical empirical potentials. GAP and MTP
established systematically improvable statistical representations of potential-energy
surfaces~\cite{bartok2010,shapeev2016}, DeePMD enabled large-scale neural-network molecular
dynamics~\cite{zeng2023}, and equivariant or message-passing architectures such as NequIP, MACE,
M3GNet, and CHGNet improved accuracy and transferability across chemical
environments~\cite{batzner2022,batatia2022,chen2022,deng2023}. Reviews emphasize their ability to
approach first-principles accuracy at substantially larger length and time scales, while also noting
domain-dependent limitations~\cite{jacobs2025}.

For defect energetics, this efficiency enables screening of many sites, reconstruction basins,
migration pathways, and system sizes that would be costly to examine exhaustively with DFT.
Universal MLIPs are therefore being explored for high-throughput defect
screening~\cite{berger2025,jacobs2025}, including the large-scale vacancy study of Berger et
al.~\cite{berger2025}. In finite GQDs, the same capability could support rapid enumeration of edge
and interior defect candidates before higher-level validation.

SevenNet is particularly relevant in this context. Park et al.\ introduced SevenNet as a scalable
GNN-based interatomic-potential package compatible with LAMMPS and based on equivariant message
passing~\cite{park2024}. Kim et al.\ later developed this platform into SevenNet-Omni, a
multi-domain universal MLIP trained on 15 open datasets and 250 million structures using a
multi-task strategy and domain-bridging sets designed for transferability across chemical domains
and even across computational protocols~\cite{kim2026}. The official SevenNet documentation
indicates that the SevenNet-Omni-i12 checkpoint is a higher-capacity version of the same training
strategy as SevenNet-Omni, effectively a higher-capacity variant of the same universal
foundation~\cite{sevennetdocs}. For the present study, this is important not only practically but
also conceptually: a model of this class claims transferability from crystals and molecules to
surfaces, defects, and finite nanostructures, and should therefore be tested specifically on locally
difficult configurations.

The limitations of universal MLIPs become apparent precisely in this regime. The 2025 study by Deng
et al.\ on systematic softening is central for understanding the failure modes of foundation and
universal MLIP models. For M3GNet, CHGNet, and MACE-MP-0, the authors found systematic
underestimation of energies and forces in tasks involving surfaces, defects, migration barriers,
phonons, and high-energy states in general, that is, configurations underrepresented in pretraining
datasets~\cite{deng2025}. They directly relate this effect to bias toward near-equilibrium
geometries and insufficient description of the curvature of the potential energy
surface~\cite{deng2025}. Independently, L\"ow et al.\ showed that even models that accurately
predict energies and forces near dynamic equilibrium can exhibit substantial errors in phonon
properties; this is another symptom of overly soft potential-energy surfaces in universal
potentials~\cite{loew2025}. The broader review by Jacobs et al.\ also emphasizes that modern MLIPs
still have serious limitations with respect to long-range interactions, magnetic states, excited
states, and case-specific out-of-distribution problems~\cite{jacobs2025}.

\subsection{Research gap and position of the present study}\label{sec:bg-gap}

Together, the literature supports three points relevant to this study. First, defect formation
energies depend on reconstruction, spin state, cell or model size, exchange-correlation treatment,
and chemical-potential convention~\cite{skowron2015,ma2009,thomas2022,kim2011}. Second, finite
nanographenes and GQDs introduce edge chemistry, passivation, radical character, and discrete
electronic states that distinguish peripheral from interior
defects~\cite{santana2013,wagner2013,vanin2010,gao2021,cho2024,martin2024,forte2008,nieman2023,shtepliuk2018}.
Third, universal MLIPs can accelerate defect screening but require domain-specific validation for
undercoordinated and out-of-distribution configurations, where systematic softening has been
reported~\cite{park2024,deng2025,berger2025,jacobs2025,kim2026,sevennetdocs,loew2025,yu2024}.

These observations motivate a controlled DFT--MLIP benchmark for coronene, \coronene, as a small
H-terminated GQD. Existing work largely treats either defect energetics in graphene and GQD models
or the development and general benchmarking of universal MLIPs. The present study focuses on a
narrower question: within coronene, is the SevenNet formation-energy error approximately uniform, or
does it vary with local chemistry, particularly undercoordination and edge-hydrogen involvement? The
comparison is intended as a system-specific benchmark rather than a universal calibration of MLIP
defect energies.

% =====================================================================
\section{Problem statement}\label{sec:problem}

\begin{figure}[tbp]
  \centering
  \includegraphics[width=0.5\textwidth]{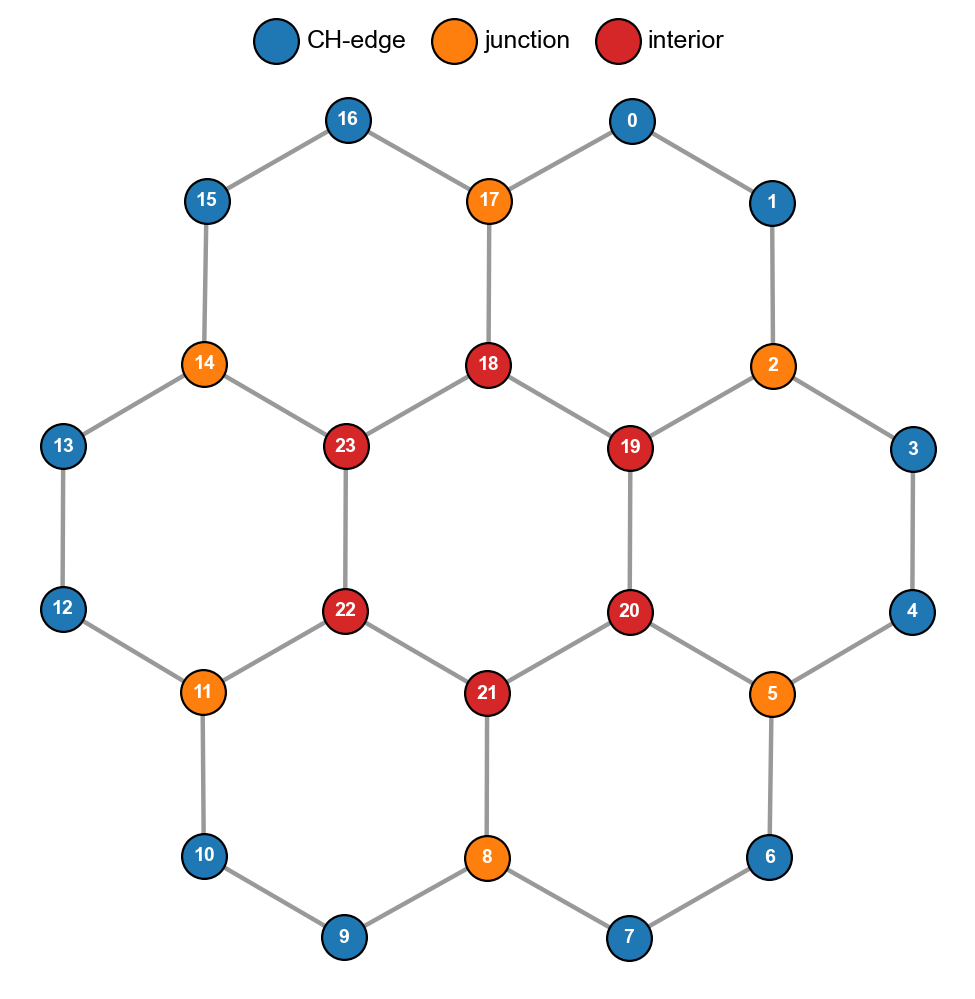}
  \caption{Coronene (\coronene) with the three symmetry-inequivalent carbon classes: twelve C--H
  edge carbons (blue), six hydrogen-free junction carbons (orange) and six interior carbons (red).
  Carbons are numbered 0--23 as in the input geometry and the energy maps. Terminating hydrogens are
  omitted for clarity.}
  \label{fig:sites}
\end{figure}

Coronene offers 24 carbon sites for the monovacancy and 30 inequivalent C--C bonds for the
divacancy and the Stone--Wales rotation. We number the carbons 0--23 after the input geometry
(Figure~\ref{fig:sites}). Under \Dsixh{} the carbons fall into three orbits:
\begin{itemize}
  \item \textbf{C--H edge carbons} (twelve: 0, 1, 3, 4, 6, 7, 9, 10, 12, 13, 15, 16): the outer rim
        atoms that each carry one terminating hydrogen;
  \item \textbf{junction carbons} (six: 2, 5, 8, 11, 14, 17): hydrogen-free rim atoms at the
        corners where two outer hexagons meet, the molecular analogue of a zigzag-edge corner site;
  \item \textbf{interior carbons} (six: 18--23): the central benzene ring, bonded only to carbon
        and reproducing in miniature the bulk-graphene environment.
\end{itemize}

The 30 bonds likewise reduce to four classes: the outer-rim C--C bonds between adjacent C--H
carbons; the edge-junction bonds; the radial ``spoke'' bonds that join a junction carbon to the
interior ring; and the inner-ring bonds. Creating a defect at each of these sites breaks a
controlled number of edge C--H bonds, from none (a fully interior divacancy) to two (removing two
adjacent hydrogen-bearing carbons), which gives a natural axis of increasing under-coordination
along which to test the surrogate. We, therefore, ask (i) how large the surrogate's formation-energy
error is, once basins are matched, and (ii) whether that error tracks the number of broken edge
bonds, with the edge-versus-interior contrast as the axis along which a bulk-trained potential is
most likely to soften.

% =====================================================================
\section{Computational methods}\label{sec:methods}

\textbf{Model and defect generation.} The pristine coronene molecule was centred in a cubic cell and
sampled at the $\Gamma$ point. Each defect was constructed from the relaxed pristine structure: a
monovacancy by deleting one carbon atom, a divacancy by deleting a bonded carbon pair, and a
Stone--Wales defect by rotating a C--C bond by \ang{90}. The input MOL connectivity was retained so
that carbon-only ring statistics could be evaluated after relaxation and the principal
reconstruction motifs identified (5-9 for monovacancies, 5-8-5 for divacancies, and 5-7-7-5 for
Stone--Wales rotations). Geometries were optimized with the BFGS algorithm in the Atomic Simulation
Environment (ASE)~\cite{larsen2017} to a maximum residual force of 0.05~\eVA. For vacancies
containing a $\sigma$-dangling bond, we also relaxed hydrogen-passivated variants in which an
additional H atom saturates the radical, allowing direct comparison with the corresponding bare
reconstruction.

\textbf{Machine-learning potential.} The surrogate is SevenNet in the 7net-omni-i12
parameterisation~\cite{park2024}: an E(3)-equivariant message-passing network pre-trained on large
first-principles datasets and used here without any system-specific fine-tuning, so the test is one
of genuine out-of-the-box transferability.

\textbf{Density functional theory.} Reference energies were computed with the
projector-augmented-wave code GPAW~\cite{enkovaara2010}, the PBE functional~\cite{perdew1996}, a
400~eV plane-wave cutoff, and $\Gamma$-point sampling. Spin polarization was enabled throughout to
accommodate open-shell vacancy states, which can carry local moments of approximately
1~$\mu_{\mathrm{B}}$~\cite{nieman2017}. A Fermi--Dirac smearing of 0.05~eV was applied, and the SCF
energy convergence threshold was $10^{-6}$~eV. Both methods used the same ASE optimization protocol;
the interatomic energy model was the principal intended difference between the two landscapes.

\textbf{Formation energies and self-consistent reservoirs.} Formation energies follow the usual
reservoir convention~\cite{skowron2015}. For a vacancy or divacancy,
\begin{equation}
  \Delta E_f = E_{\mathrm{def}} - E_0 + n_{\mathrm{C}}\,\mu_{\mathrm{C}} + n_{\mathrm{H}}\,\mu_{\mathrm{H}},
  \label{eq:Ef-vac}
\end{equation}
where $E_0$ and $E_{\mathrm{def}}$ are the relaxed energies of pristine and defective coronene, and
$n_{\mathrm{C}}$ and $n_{\mathrm{H}}$ count the carbon and hydrogen atoms removed (negative when a
passivating hydrogen is added). The carbon reservoir $\mu_{\mathrm{C}}$ is the energy per atom of a
relaxed graphene sheet and the hydrogen reservoir is $\mu_{\mathrm{H}} = \tfrac{1}{2}E(\ce{H2})$;
each chemical potential is recomputed with the same calculator that produced the structure energies.
For DFT this gives $E_0 = -259.975$~eV, $\mu_{\mathrm{C}} = -8.861$~eV and
$\mu_{\mathrm{H}} = -3.355$~eV; for the MLIP, $E_0 = -262.160$~eV, $\mu_{\mathrm{C}} = -9.221$~eV and
$\mu_{\mathrm{H}} = -3.364$~eV. Because each engine uses its own self-consistent reservoir and, in
the paired comparison below, is scored on the \emph{same} relaxed geometry, the difference between
the two formation energies isolates the energy model and nothing else. A Stone--Wales rotation
removes no atoms, so
\begin{equation}
  \Delta E_f = E_{\mathrm{SW}} - E_0 .
  \label{eq:Ef-sw}
\end{equation}

\textbf{Two-track comparison.} We report two complementary datasets. The automatic landscape
constructs a defect at every carbon site or C--C bond and relaxes each structure independently with
both methods. It provides a symmetry-resolved map over all 24 sites and 30 bonds; the within-class
formation-energy spread (maximum minus minimum), which should vanish for symmetry-equivalent sites
that reach the same basin, serves as a diagnostic of basin-sensitive relaxations. The paired
recalculation set contains 28 relaxed vacancy and divacancy geometries, each evaluated with both
energy models without changing the structure. The resulting DFT--MLIP difference is therefore a
basin-matched measure of the surrogate bias.

% =====================================================================
\section{Results and discussion}\label{sec:results}

We discuss the defects in order of increasing under-coordination, from coordination-preserving
Stone--Wales rotations through divacancies to radical-bearing monovacancies, and analyze the
surrogate error as a function of broken edge bonds. Table~\ref{tab:paired} summarizes the paired
formation-energy means and biases.

\begin{table}[tbp]
  \caption{Paired, basin-matched formation energies of coronene point defects: each reconstruction
  is relaxed once and scored by both engines. ``Broken C--H'' is the number of edge C--H bonds
  severed and left unsaturated by the reconstruction. For each group we give the mean \Ef{} (eV)
  from DFT and from the MLIP, the mean bias
  $\delta = \Delta E_f(\mathrm{MLIP}) - \Delta E_f(\mathrm{DFT})$, and the number $n$ of
  configurations. Stone--Wales conserves atoms and coordination and is taken from the full landscape
  (30 bonds). All 28 vacancy and divacancy biases are negative, indicating systematic MLIP
  underestimation within the paired dataset.}
  \label{tab:paired}
  \centering
  \small
  \begin{tabular}{@{}l c S[table-format=2.0] S[table-format=1.3] S[table-format=1.3]
                  S[table-format=-1.3]@{}}
    \toprule
    Defect & {Broken C--H} & {$n$} & {DFT mean} & {MLIP mean} & {$\delta = \text{MLIP} - \text{DFT}$} \\
    \midrule
    Stone--Wales & 0 & 30 & 1.891 & 1.878 & -0.013 \\
    \addlinespace[2pt]
    Monovacancy  & 0 &  2 & 1.808 & 1.509 & -0.299 \\
    Monovacancy  & 1 &  4 & 2.497 & 1.758 & -0.739 \\
    \addlinespace[2pt]
    Divacancy    & 0 & 17 & 1.139 & 0.487 & -0.653 \\
    Divacancy    & 1 &  2 & 4.593 & 3.737 & -0.856 \\
    Divacancy    & 2 &  3 & 5.798 & 4.686 & -1.112 \\
    \bottomrule
  \end{tabular}
\end{table}

\subsection{Stone--Wales: coordination-conserving, and the surrogate is exact}\label{sec:sw}

Stone--Wales rotations provide a coordination-preserving test case because they conserve atom count
and carbon coordination while primarily distorting bond angles. Across all 30 bonds, the mean
formation energies are 1.891~eV (DFT) and 1.878~eV (MLIP), a difference of 0.013~eV. The class
ordering is also preserved: both methods place outer-rim rotations lowest (approximately
1.48--1.57~eV) and inner-ring rotations highest (approximately 2.41--2.54~eV), while the central
rotation relaxes to a 5-7-7-5 motif. SevenNet therefore reproduces the sampled Stone--Wales
landscape accurately; this result should not be extrapolated to under-coordinated vacancy defects.

\subsection{The softening bias scales with under-coordination}\label{sec:bias}

\begin{figure}[tbp]
  \centering
  \includegraphics[width=\textwidth]{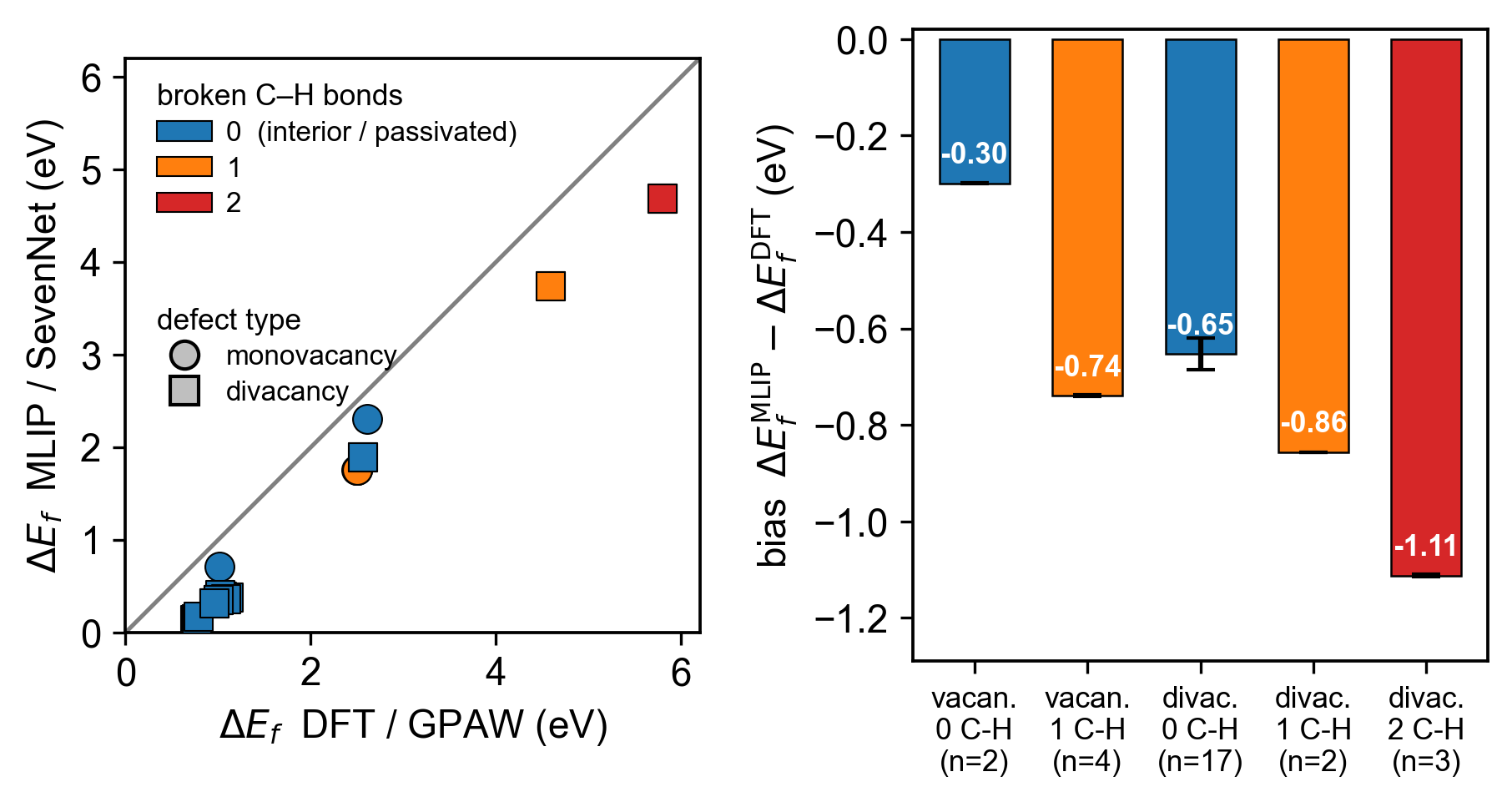}\\[-1pt]
  \makebox[\textwidth][l]{%
    \hspace{0.273\textwidth}\makebox[0pt]{\small(a)}%
    \hspace{0.522\textwidth}\makebox[0pt]{\small(b)}}
  \caption{Basin-matched comparison of MLIP and DFT formation energies for the 28 paired
  configurations. (a)~Parity plot: every MLIP value lies below the diagonal; color indicates the
  number of broken edge C--H bonds, and marker shape indicates defect type. (b)~Mean formation-energy
  bias (MLIP~$-$~DFT), grouped by defect class and number of broken edge C--H bonds. Within both the
  monovacancy and divacancy series, the bias becomes more negative as the number of broken bonds
  increases. Error bars denote the within-group standard deviation and describe dispersion among
  configurations rather than uncertainty from repeated measurements.}
  \label{fig:softening}
\end{figure}

Vacancies show a different regime. For all 28 paired vacancy and divacancy configurations, the MLIP
formation energy lies below the DFT value (Figure~\ref{fig:softening}a). This consistently negative
bias is direct evidence of systematic softening in the present dataset and agrees with the expected
effect of training predominantly on near-equilibrium structures~\cite{deng2025}.

The magnitude of underestimation is not constant and increases with under-coordination within each
defect family (Figure~\ref{fig:softening}b and Table~\ref{tab:paired}). For monovacancies, the mean
deficit rises from 0.30~eV without a broken edge C--H bond to 0.74~eV with one broken bond. For
divacancies, it increases from 0.65~eV with no unsaturated edge bond to 0.86~eV with one and
1.11~eV with two. The increments therefore depend on defect type; the number of broken edge bonds is
a useful low-dimensional descriptor, but not a universal linear correction. The comparatively small
within-group spreads support an association between local coordination and bias over the sampled
configurations.

This trend has a direct consequence for calibration. A single additive correction is not
appropriate because the required correction spans 0.30--1.11~eV and differs between monovacancies
and divacancies. A practical calibration should therefore include both defect class and local
under-coordination; the number of broken edge C--H bonds provides a simple structural descriptor.

\subsection{Hydrogen trapping controls ring closure and reconstruction basins}\label{sec:htrap}

\begin{figure}[tbp]
  \centering
  \includegraphics[width=0.78\textwidth]{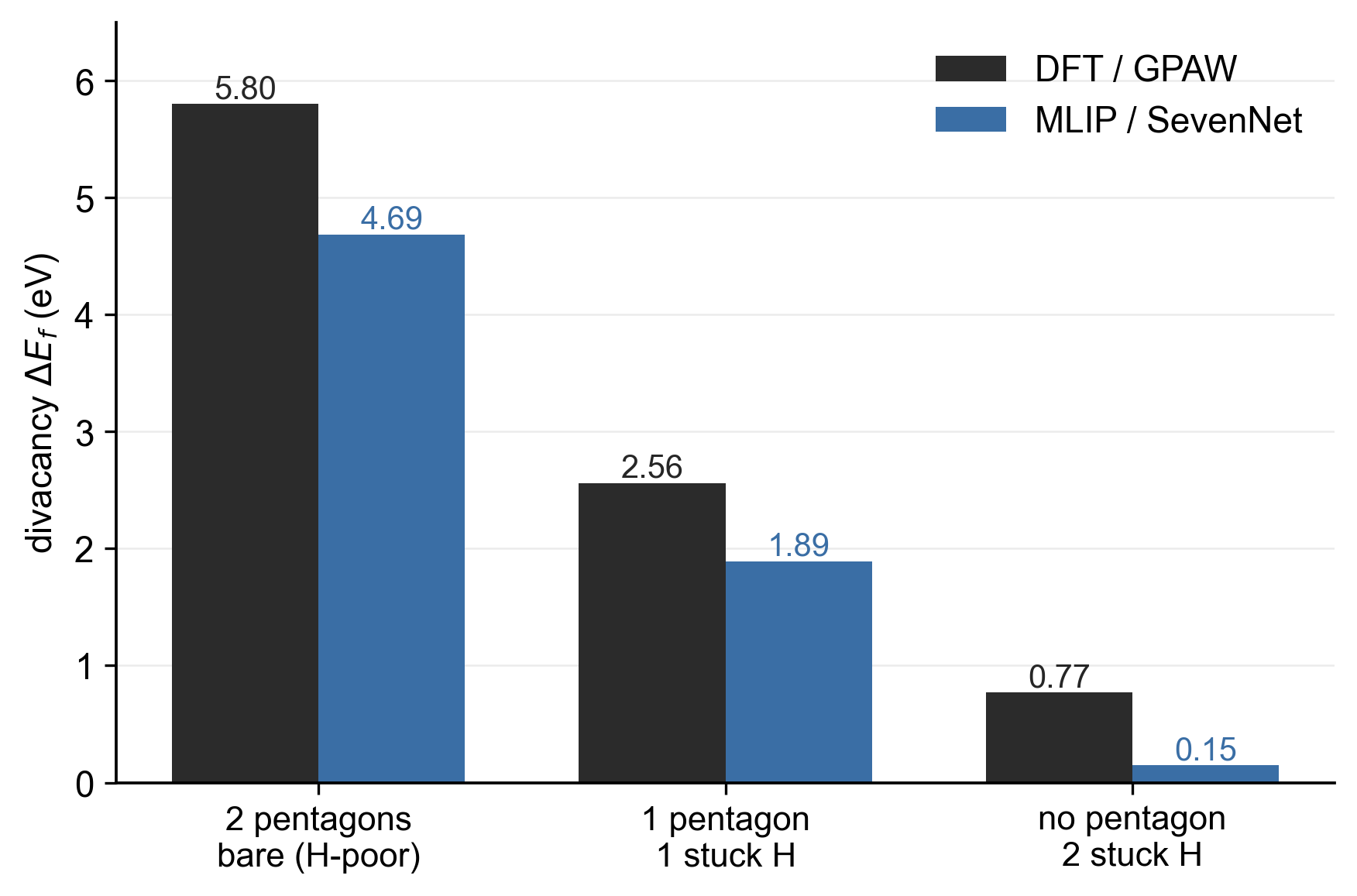}
  \caption{Formation energies of three hydrogen-dependent reconstruction basins for the same
  edge--junction corner divacancy, calculated with DFT/GPAW (black) and MLIP/SevenNet (blue). The
  hydrogen-poor defect forms two pentagons; trapping one hydrogen leaves one pentagon; trapping two
  hydrogens suppresses pentagon formation. Both methods give the same basin ordering, whereas the
  MLIP underestimates every absolute formation energy.}
  \label{fig:reconstruction}
\end{figure}

A nominal edge divacancy can relax into several reconstruction basins depending on hydrogen
trapping. Removing adjacent edge carbons leaves two undercoordinated carbons facing each other
across the void. In the hydrogen-poor state, they rebond and form a bare 5-5 pair of pentagons, with
a DFT formation energy of 5.80~eV. Trapping one hydrogen blocks one closure pathway and yields a
single-pentagon state at 2.56~eV. Trapping two hydrogen atoms suppresses pentagon formation and
lowers the DFT formation energy to 0.77~eV (Figure~\ref{fig:reconstruction}). Hydrogen termination
therefore controls both the carbon-ring topology and the energy of the resulting basin.

The formation energy decreases stepwise as hydrogen trapping suppresses ring closure, although the
spacing between the levels is not uniform. The DFT energy decreases by 3.24~eV between the
two-pentagon and one-pentagon states and by 1.79~eV between the one-pentagon and no-pentagon states.
The corresponding MLIP decreases are 2.80 and 1.74~eV. These structures are distinct relaxed
minima. The basin reached by an unconstrained optimization can nevertheless depend on hydrogen
placement during relaxation, contributing to the scatter in the automatic landscape discussed in the
next section.

SevenNet preserves the ordering of the three reconstruction basins in this test
(Figure~\ref{fig:reconstruction}). The MLIP formation energies are 4.69, 1.89, and 0.15~eV for the
two-pentagon, one-pentagon, and no-pentagon states, respectively, compared with 5.80, 2.56, and
0.77~eV from DFT. Thus, the MLIP underestimates each absolute formation energy but retains the
relative stability sequence for these three basins. This supports its use for preliminary ranking
of such reconstruction candidates, provided that final formation energies are validated with DFT.

\subsection{Symmetry consistency as a free flag for the fragile sites}\label{sec:symmetry}

\begin{figure}[tbp]
  \centering
  \includegraphics[width=0.88\textwidth]{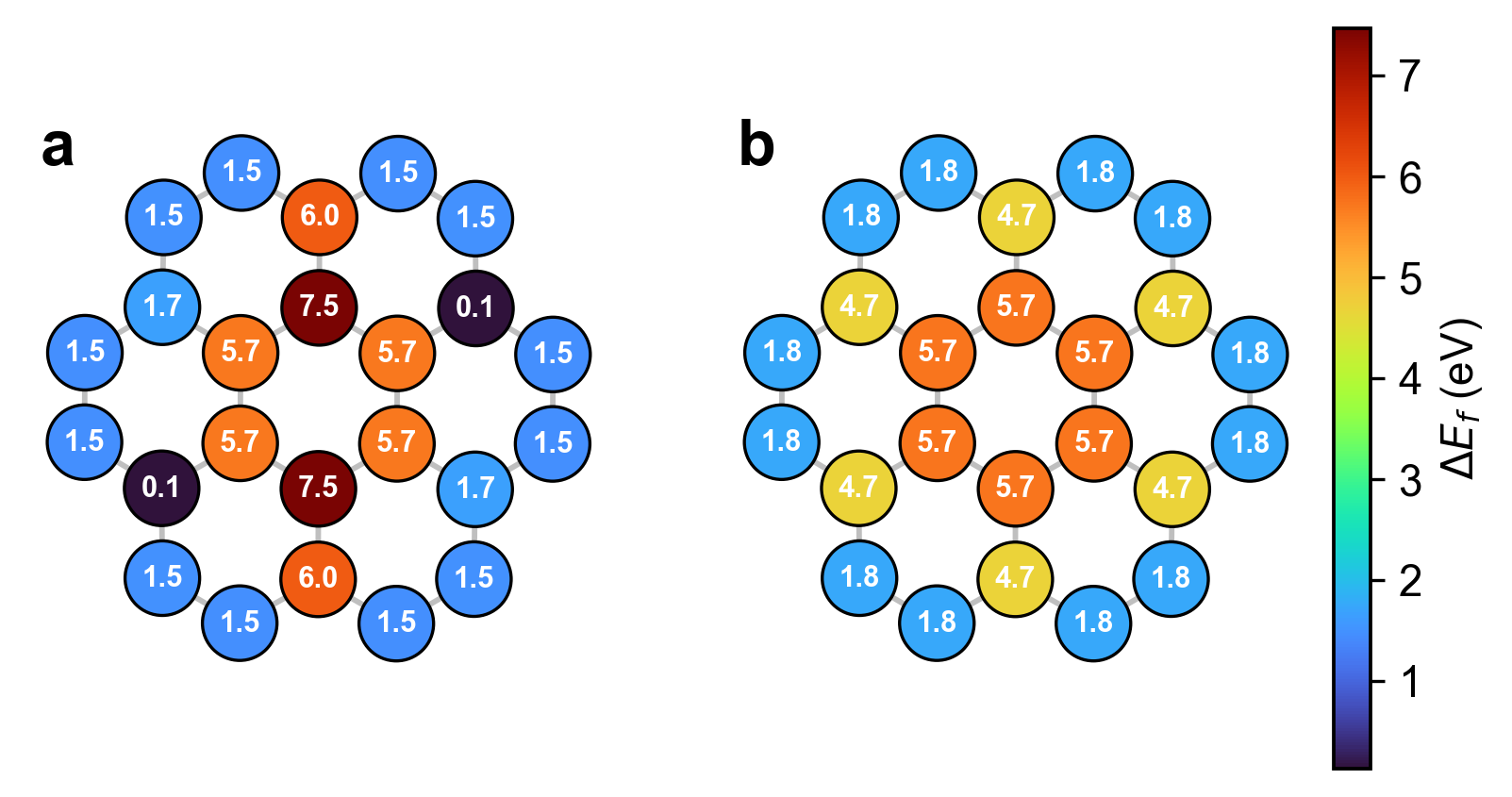}
  \caption{Monovacancy formation-energy maps of coronene from (a)~spin-polarized DFT/GPAW and
  (b)~the MLIP/SevenNet surrogate; each carbon is colored and labeled by its formation energy (eV)
  from the automatic, independently relaxed landscape. The surrogate respects the molecule's sixfold
  symmetry, whereas independent DFT relaxations of the equivalent junction carbons span 0.1--6.0~eV
  because they converge to different reconstruction basins; the corresponding within-class spread
  is 5.85~eV, although symmetry requires zero.}
  \label{fig:maps}
\end{figure}

The paired comparison of Section~\ref{sec:bias} requires both methods to be scored on the same
relaxed structure, which raises the practical question of how an automated workflow can detect cases
in which independent relaxations reach different basins. Molecular symmetry provides a diagnostic
without additional reference calculations. Under \Dsixh{} symmetry, equivalent sites should have the
same formation energy if they converge to the same basin. Figure~\ref{fig:maps} shows the two
monovacancy maps. SevenNet returns one value per class to three decimal places (within-class spread
$\leq 0.002$~eV), consistent with the six-fold symmetry to numerical precision. Spin-polarized DFT
reproduces this behavior for the closed-shell classes, whereas independently relaxed junction
monovacancies span 5.85~eV (0.13 to 5.98~eV) and the interior class spans 1.76~eV.

Because these sites are symmetry equivalent, the large DFT spread is attributable to independent
optimizations reaching different reconstruction basins rather than to physical lifting of the
degeneracy. Junction carbons resemble zigzag-edge corners, which can undergo several competing
reconstructions~\cite{koskinen2008,rodrigues2011,kroes2011,kim2013,jiang2025}. In the present
dataset, the resulting within-class spread identifies the open-shell junction and interior
monovacancy classes as basin-sensitive, while the closed-shell classes remain symmetry consistent.
This suggests a practical screening heuristic: classes with small symmetry spread may be retained
for surrogate-based ranking, whereas classes with large spread can be escalated to
symmetry-resolved, reconstruction-by-reconstruction electronic-structure calculations. The
threshold and any energy adjustment would require validation on broader datasets before use as a
general decision rule.

% =====================================================================
\section{Conclusions}\label{sec:conclusions}

Benchmarking SevenNet (7net-omni-i12) against spin-polarized PBE DFT (GPAW) for point defects in
coronene provides a system-specific, defect-resolved assessment of surrogate softening under matched
geometries and self-consistent chemical-potential reservoirs. For the Stone--Wales rotations sampled
here, the mean DFT--MLIP difference is 0.013~eV. For the 28 paired vacancy and divacancy
configurations, SevenNet underestimates every formation energy, with mean deficits increasing from
0.30 to 0.74~eV for monovacancies and from 0.65 to 1.11~eV for divacancies as the number of broken
edge C--H bonds increases. These observations show that a single additive offset would not describe
the present data adequately. They identify local under-coordination as a promising explanatory
descriptor, but do not by themselves establish a validated predictive correction.

SevenNet also preserves the ordering of the three hydrogen-dependent reconstruction basins examined
here and reproduces symmetry-equivalent sites to within a few meV. In this system, the within-class
symmetry spread therefore acts as a useful indicator of basin-sensitive relaxations: large spreads
mark classes for which independently optimized structures should be examined more carefully. These
findings suggest a staged workflow in which MLIP screening is used for candidate generation and
preliminary ranking, followed by DFT validation for absolute vacancy energies and multi-basin
configurations. An under-coordination-dependent energy adjustment may be worth investigating, but
its form, threshold values, and transferability across GQD sizes, edge terminations, defect classes,
and MLIP architectures remain to be established on larger independent datasets.

Applications of MLIPs to orientational defects in polymers~\cite{kiv2025} indicate that
surrogate-assisted defect modeling extends beyond graphene. The present coronene results therefore
suggest, rather than establish, a potentially transferable validation strategy based on chemically
interpretable coordination descriptors and symmetry-consistency checks. Testing that strategy across
broader material and defect datasets is a natural direction for future work.

% =====================================================================
\section*{Acknowledgements}
\addcontentsline{toc}{section}{Acknowledgements}

Computations were performed on the local cluster of the South Ukrainian National Pedagogical
University named after K.\,D.~Ushynsky. The authors thank the developers of ASE, GPAW and SevenNet
for their open-source tools.

\section*{Declaration on Generative AI}
\addcontentsline{toc}{section}{Declaration on Generative AI}

The authors used a generative-AI assistant for code scaffolding of the analysis scripts and for
language editing. After using these tools, the authors reviewed and edited the content as needed and
take full responsibility for the publication's content.

% =====================================================================


\begin{thebibliography}{48}

\bibitem{geng2026}
Z.~L. Geng, et al., Recent advances on graphene quantum dots and their derivatives for energy
storage devices, Int. J. Extrem. Manuf. 8 (2026) 022005. \doi{10.1088/2631-7990/ae2221}.

\bibitem{batzner2022}
S.~Batzner, et al., E(3)-equivariant graph neural networks for data-efficient and accurate
interatomic potentials, Nat. Commun. 13 (2022) 2453. \doi{10.1038/s41467-022-29939-5}.

\bibitem{batatia2022}
I.~Batatia, D.~P. Kov\'acs, G.~N.~C. Simm, C.~Ortner, G.~Cs\'anyi, MACE: higher order equivariant
message passing neural networks for fast and accurate force fields, Adv. Neural Inf. Process. Syst.
35 (2022) 11423--11436. \doi{10.52202/068431-0830}.

\bibitem{park2024}
Y.~Park, J.~Kim, S.~Hwang, S.~Han, Scalable parallel algorithm for graph neural network interatomic
potentials in molecular dynamics simulations, J. Chem. Theory Comput. 20 (2024) 4857--4868.
\doi{10.1021/acs.jctc.4c00190}.

\bibitem{deng2025}
B.~Deng, et al., Systematic softening in universal machine learning interatomic potentials, npj
Comput. Mater. 11 (2025) 9. \doi{10.1038/s41524-024-01500-6}.

\bibitem{nieman2017}
R.~Nieman, A.~Das, A.~J.~A. Aquino, R.~G. Amorim, F.~B.~C. Machado, H.~Lischka, Single and double
carbon vacancies in pyrene as first models for graphene defects: a survey of the chemical reactivity
toward hydrogen, Chem. Phys. 482 (2017) 346--354. \doi{10.1016/j.chemphys.2016.08.007}.

\bibitem{skowron2015}
S.~T. Skowron, I.~V. Lebedeva, A.~M. Popov, E.~Bichoutskaia, Energetics of atomic scale structure
changes in graphene, Chem. Soc. Rev. 44 (2015) 3143--3176. \doi{10.1039/C4CS00499J}.

\bibitem{ma2004}
Y.~Ma, P.~O. Lehtinen, A.~S. Foster, R.~M. Nieminen, Magnetic properties of vacancies in graphene
and single-walled carbon nanotubes, New J. Phys. 6 (2004) 68. \doi{10.1088/1367-2630/6/1/068}.

\bibitem{lee2005}
G.-D. Lee, C.~Z. Wang, E.~Yoon, N.~M. Hwang, D.~Y. Kim, K.~M. Ho, Diffusion, coalescence, and
reconstruction of vacancy defects in graphene layers, Phys. Rev. Lett. 95 (2005) 205501.
\doi{10.1103/PhysRevLett.95.205501}.

\bibitem{ma2009}
J.~Ma, D.~Alf\`e, A.~Michaelides, E.~Wang, Stone--Wales defects in graphene and other planar
sp$^2$-bonded materials, Phys. Rev. B 80 (2009) 033407. \doi{10.1103/PhysRevB.80.033407}.

\bibitem{tiwari2023}
S.~K. Tiwari, et al., Stone--Wales defect in graphene, Small 19 (2023) 2303340.
\doi{10.1002/smll.202303340}.

\bibitem{santana2013}
A.~Santana, A.~M. Popov, E.~Bichoutskaia, Stability and dynamics of vacancy in graphene flakes: edge
effects, Chem. Phys. Lett. 557 (2013) 80--87. \doi{10.1016/j.cplett.2012.11.077}.

\bibitem{wagner2013}
P.~Wagner, et al., Stable hydrogenated graphene edge types: normal and reconstructed Klein edges,
Phys. Rev. B 88 (2013) 094106. \doi{10.1103/PhysRevB.88.094106}.

\bibitem{vanin2010}
M.~Vanin, J.~Gath, K.~S. Thygesen, K.~W. Jacobsen, First-principles calculations of graphene
nanoribbons in gaseous environments: structural and electronic properties, Phys. Rev. B 82 (2010)
195411. \doi{10.1103/PhysRevB.82.195411}.

\bibitem{gao2021}
Y.~Gao, D.~Xu, T.~Cui, D.~Li, Stability of hydrogen-terminated graphene edges, Phys. Chem. Chem.
Phys. 23 (2021) 13261--13266. \doi{10.1039/D1CP01384J}.

\bibitem{cho2024}
H.~Cho, G.~Bae, B.~H. Hong, Engineering functionalization and properties of graphene quantum dots
with controllable synthesis for energy and display applications, Nanoscale 16 (2024) 3347--3378.
\doi{10.1039/D3NR05842E}.

\bibitem{martin2024}
A.~Mart\'{\i}n Santa Dar\'{\i}a, L.~Gonz\'alez-S\'anchez, S.~G\'omez, Coronene: a model for
ultrafast dynamics in graphene nanoflakes and PAHs, Phys. Chem. Chem. Phys. 26 (2024) 174--184.
\doi{10.1039/D3CP03656A}.

\bibitem{forte2008}
G.~Forte, A.~Grassi, G.~M. Lombardo, A.~La~Magna, G.~G.~N. Angilella, R.~Pucci, et al., Modeling
vacancies and hydrogen impurities in graphene: a molecular point of view, Phys. Lett. A 372 (2008)
6168--6174. \doi{10.1016/j.physleta.2008.08.014}.

\bibitem{thomas2022}
D.~M. Thomas, Y.~Asiri, N.~D. Drummond, Point defect formation energies in graphene from diffusion
quantum Monte Carlo and density functional theory, Phys. Rev. B 105 (2022) 184114.
\doi{10.1103/PhysRevB.105.184114}.

\bibitem{machado2015}
F.~B.~C. Machado, A.~J.~A. Aquino, H.~Lischka, The electronic states of a double carbon vacancy
defect in pyrene: a model study for graphene, Phys. Chem. Chem. Phys. 17 (2015) 12778--12785.
\doi{10.1039/C4CP05751A}.

\bibitem{pinheiro2019}
M.~Pinheiro~Jr., D.~V.~V. Cardoso, A.~J.~A. Aquino, F.~B.~C. Machado, H.~Lischka, The
characterization of electronic defect states of single and double carbon vacancies in graphene
sheets using molecular density functional theory, Mol. Phys. 117 (2019) 1519--1531.
\doi{10.1080/00268976.2019.1567848}.

\bibitem{nieman2023}
R.~Nieman, V.~P. Oliveira, B.~Jayee, A.~J.~A. Aquino, F.~B.~C. Machado, H.~Lischka, High-level
multireference investigations on the electronic states in single-vacancy graphene defects using a
pyrene-SV model, J. Phys. Chem. A 127 (2023) 8287--8296. \doi{10.1021/acs.jpca.3c04099}.

\bibitem{shtepliuk2018}
I.~Shtepliuk, R.~Yakimova, Interband transitions in closed-shell vacancy-containing graphene
quantum dots complexed with heavy metals, Phys. Chem. Chem. Phys. 20 (2018) 21528--21543.
\doi{10.1039/C8CP03306D}.

\bibitem{lin2015}
C.-K. Lin, Theoretical study of graphene nanoflakes with nitrogen and vacancy defects: effects of
flake sizes and defect types on electronic excitation properties, J. Phys. Chem. C 119 (2015)
27131--27144. \doi{10.1021/acs.jpcc.5b08318}.

\bibitem{valencia2017}
A.~M. Valencia, M.~J. Caldas, Single vacancy defect in graphene: insights into its magnetic
properties, Phys. Rev. B 96 (2017) 125431. \doi{10.1103/PhysRevB.96.125431}.

\bibitem{kim2011}
Y.~Kim, J.~Ihm, E.~Yoon, G.-D. Lee, Dynamics and stability of divacancy defects in graphene, Phys.
Rev. B 84 (2011) 075445. \doi{10.1103/PhysRevB.84.075445}.

\bibitem{amorim2007}
R.~G. Amorim, A.~Fazzio, A.~Antonelli, F.~D. Novaes, A.~J.~R. da~Silva, Divacancies in graphene and
carbon nanotubes, Nano Lett. 7 (2007) 2459--2462. \doi{10.1021/nl071217v}.

\bibitem{lusk2008}
M.~T. Lusk, L.~D. Carr, Nanoengineering defect structures on graphene, Phys. Rev. Lett. 100 (2008)
175503. \doi{10.1103/PhysRevLett.100.175503}.

\bibitem{berger2025}
E.~Berger, M.~Bagheri, H.-P. Komsa, Screening of material defects using universal machine-learning
interatomic potentials, Small 21 (2025) e03956. \doi{10.1002/smll.202503956}.

\bibitem{koskinen2008}
P.~Koskinen, S.~Malola, H.~H\"akkinen, Self-passivating edge reconstructions of graphene, Phys. Rev.
Lett. 101 (2008) 115502. \doi{10.1103/PhysRevLett.101.115502}.

\bibitem{rodrigues2011}
J.~N.~B. Rodrigues, P.~A.~D. Gon\c{c}alves, N.~F.~G. Rodrigues, R.~M. Ribeiro, J.~M.~B. Lopes dos
Santos, N.~M.~R. Peres, Zigzag graphene nanoribbon edge reconstruction with Stone--Wales defects,
Phys. Rev. B 84 (2011) 155435. \doi{10.1103/PhysRevB.84.155435}.

\bibitem{kroes2011}
J.~M.~H. Kroes, M.~A. Akhukov, J.~H. Los, N.~Pineau, A.~Fasolino, Mechanism and free-energy barrier
of the type-57 reconstruction of the zigzag edge of graphene, Phys. Rev. B 83 (2011) 165411.
\doi{10.1103/PhysRevB.83.165411}.

\bibitem{kim2013}
K.~Kim, et al., Atomically perfect torn graphene edges and their reversible reconstruction, Nat.
Commun. 4 (2013) 2723. \doi{10.1038/ncomms3723}.

\bibitem{bartok2010}
A.~P. Bart\'ok, M.~C. Payne, R.~Kondor, G.~Cs\'anyi, Gaussian approximation potentials: the
accuracy of quantum mechanics, without the electrons, Phys. Rev. Lett. 104 (2010) 136403.
\doi{10.1103/PhysRevLett.104.136403}.

\bibitem{shapeev2016}
A.~V. Shapeev, Moment tensor potentials: a class of systematically improvable interatomic
potentials, Multiscale Model. Simul. 14 (2016) 1153--1173. \doi{10.1137/15M1054183}.

\bibitem{zeng2023}
J.~Zeng, et al., DeePMD-kit v2: a software package for deep potential models, J. Chem. Phys. 159
(2023) 054801. \doi{10.1063/5.0155600}.

\bibitem{chen2022}
C.~Chen, S.~P. Ong, A universal graph deep learning interatomic potential for the periodic table,
Nat. Comput. Sci. 2 (2022) 718--728. \doi{10.1038/s43588-022-00349-3}.

\bibitem{deng2023}
B.~Deng, et al., CHGNet as a pretrained universal neural network potential for charge-informed
atomistic modelling, Nat. Mach. Intell. 5 (2023) 1031--1041. \doi{10.1038/s42256-023-00716-3}.

\bibitem{jacobs2025}
R.~Jacobs, et al., A practical guide to machine learning interatomic potentials -- status and
future, Curr. Opin. Solid State Mater. Sci. 35 (2025) 101214. \doi{10.1016/j.cossms.2025.101214}.

\bibitem{kim2026}
J.~Kim, et al., Optimizing cross-domain transfer for universal machine learning interatomic
potentials, Nat. Commun. 17 (2026) 3432. \doi{10.1038/s41467-026-70195-8}.

\bibitem{sevennetdocs}
SevenNet 0.13.0 documentation, Pretrained models, 2026. URL:
\url{https://sevennet.readthedocs.io/en/latest/user_guide/pretrained.html} (accessed 2026-07-02).

\bibitem{loew2025}
A.~Loew, D.~Sun, H.-C. Wang, S.~Botti, M.~A.~L. Marques, Universal machine learning interatomic
potentials are ready for phonons, npj Comput. Mater. 11 (2025) 178.
\doi{10.1038/s41524-025-01650-1}.

\bibitem{yu2024}
H.~Yu, M.~Giantomassi, G.~Materzanini, J.~Wang, G.-M. Rignanese, Systematic assessment of various
universal machine-learning interatomic potentials, MGE Adv. 2 (2024) e58. \doi{10.1002/mgea.58}.

\bibitem{larsen2017}
A.~H. Larsen, et al., The atomic simulation environment: a Python library for working with atoms, J.
Phys.: Condens. Matter 29 (2017) 273002. \doi{10.1088/1361-648X/aa680e}.

\bibitem{enkovaara2010}
J.~Enkovaara, et al., Electronic structure calculations with GPAW: a real-space implementation of
the projector augmented-wave method, J. Phys.: Condens. Matter 22 (2010) 253202.
\doi{10.1088/0953-8984/22/25/253202}.

\bibitem{perdew1996}
J.~P. Perdew, K.~Burke, M.~Ernzerhof, Generalized gradient approximation made simple, Phys. Rev.
Lett. 77 (1996) 3865--3868. \doi{10.1103/PhysRevLett.77.3865}.

\bibitem{jiang2025}
W.~Jiang, Z.-F. Liu, Q.-P. Wu, X.~Xiao, Role of edge reconfiguration in generating corner states in
zigzag graphene nanoribbons, Physica E 172 (2025) 116260. \doi{10.1016/j.physe.2025.116260}.

\bibitem{kiv2025}
A.~Kiv, et al., Machine learning interatomic potentials and orientational defects in polymers, in:
P.~Petkov, M.~E. Achour, C.~Popov (Eds.), Nanotechnological Advances in Environmental, Cyber and
CBRN Security, NATO Science for Peace and Security Series B: Physics and Biophysics, Springer,
Dordrecht, 2025, pp. 91--104. \doi{10.1007/978-94-024-2316-7_6}.

\end{thebibliography}
\end{document}